\documentclass[conference]{IEEEtran}
\IEEEoverridecommandlockouts
\usepackage{cite}
\usepackage{amsmath,amssymb,amsfonts}
\usepackage{algorithmic}
\usepackage{graphicx}
\usepackage{textcomp}
\usepackage{xcolor}
\usepackage{url}
\usepackage{enumitem}
\usepackage{dsfont}
\usepackage{subfigure}

\def \bs           {\boldsymbol{s}}

\def \bu           {\boldsymbol{u}}

\def \bx           {\boldsymbol{x}}
\def \by           {\boldsymbol{y}}
\def \bz           {\boldsymbol{z}}

\def \bA           {\boldsymbol{A}}
\def \bB           {\boldsymbol{B}}

\def \bD           {\boldsymbol{D}}

\def \bI           {\boldsymbol{I}}
\def \bJ           {\boldsymbol{J}}

\def \bL           {\boldsymbol{L}}

\def \bS           {\boldsymbol{S}}

\def \bW           {\boldsymbol{W}}

\def \bY           {\boldsymbol{Y}}

\def \balpha       {\boldsymbol{\alpha}}

\def \bGamma       {\boldsymbol{\Gamma}}

\def \bepsilon     {\boldsymbol{\epsilon}}

\def \bUpsilon     {\boldsymbol{\Upsilon}}

\def \bchi         {\boldsymbol{\chi}}

\def \bOmega       {\boldsymbol{\Omega}}

\def \complexC     {\mathbb{C}}

\def \calG         {\mathcal{G}}

\def \arg          {\mathrm{arg}}

\def \st           {\mathrm{s.t.}}

\def\BibTeX{{\rm B\kern-.05em{\sc i\kern-.025em b}\kern-.08em
    T\kern-.1667em\lower.7ex\hbox{E}\kern-.125emX}}

\begin{document}

\title{Deep Radar Waveform Design for Efficient Automotive Radar Sensing
}

\author{\IEEEauthorblockN{Shahin Khobahi\IEEEauthorrefmark{1}, Arindam Bose\IEEEauthorrefmark{2}, and Mojtaba Soltanalian\IEEEauthorrefmark{3}}\\
\IEEEauthorblockA{\textit{Department of Electrical and Computer Engineering, University of Illinois at Chicago,
Chicago, IL 60607, USA} \\
Email: \{\IEEEauthorrefmark{1}skhoba2, \IEEEauthorrefmark{2}abose4, \IEEEauthorrefmark{3}msol\}@uic.edu}\vspace{-10pt}
\thanks{This work was supported in part by the U.S. National Science Foundation Grant ECCS-1809225, and a Discovery Partners Institute (DPI) Seed Award. The first two authors contributed equally to this work. Corresponding author: (e-mail: abose4@uic.edu).}
}
\maketitle

\begin{abstract}
	In radar systems, unimodular (or constant-modulus) waveform design plays an important role in achieving better clutter/interference rejection, as well as a more accurate estimation of the target parameters. The design of such sequences has been studied widely in the last few decades, with most design algorithms requiring sophisticated \textit{a priori} knowledge of environmental parameters which may be difficult to obtain in real-time scenarios. In this paper, we propose a novel hybrid model-driven and data-driven architecture that adapts to the ever changing environment and allows for adaptive unimodular waveform design. In particular, the approach lays the groundwork for developing extremely low-cost waveform design and processing frameworks for radar systems deployed in autonomous vehicles. The proposed model-based deep architecture imitates a well-known unimodular signal design algorithm in its structure, and can quickly infer statistical information from the environment using the observed data. Our numerical experiments portray the advantages of using the proposed method for efficient radar waveform design in time-varying environments.
\end{abstract}

\section{Introduction}\label{sec:intro}
Waveform design for active sensing has been of interest to engineers, system theorists and mathematicians in the last sixty years. In the last decade, however, civilian radar applications such as the use of radar in autonomous cars have attracted much-deserved attention towards enhanced resolvability for advanced safety. 
In vehicular applications, the radar technology offers excellent resolvability and immunity to bad weather conditions in comparison to visible and infrared imaging techniques. However, the cost overheads of ultra-high frequency radar signal processors is excessive, which limits a mass deployment of radar-based advanced vehicular safety features. Reliable and low-cost deep learning-based algorithms and hardware may promise a solution to such difficulties.

The quality of automotive radar sensors' measurements depends strongly on the transmit waveform design process \cite{Hermann2012}. There exist several approaches to tackle the task of waveform design in such radar systems \cite{naghsh2014radar,  wddbook, mimobook, skolnik, demaioclutter, demaioclutter2, naghsh2013unified, Naghsh_Dopplerrobust, petremimorangecompress, iGLRT, hu2017locating, bellMI, paperfred, farina, blum, petremimo2, Delong1967, ameri2019one, spafford1967, soltanalian2012computational, tang, mimojammer, kaysimo, 4644058, soltanalian2014designing, 6472022, bose2018constructing, bose2019joint, 8645383, bose2019waveform}, which rely on known radar models. In such model-based approaches\nocite{984612, 4721121, 6978884}, one only considers a simplified mathematical model and often do not take into account the intricate interactions innate to the kind of complex information systems that are common in real world. On the other hand, in a purely data-driven approach, including deep learning techniques, one do not need an explicit mathematical model of the problem, and should be able to  use  the available data at hand for designing the waveforms. The major shortcoming of the data-driven approach stems from the fact that it is unclear how to incorporate the existing knowledge of the system model in the processing stage. Namely, purely data-driven approaches have a wider applicability at the cost of interpretability, and in some cases, reliability  \cite{7944481, khobahi2019model}.  In this paper, we seek to bridge the gap between the model-based and data-driven approaches, and propose a novel methodology in order to design efficient waveforms for automotive radars by making use of the \textit{deep unfolding framework} \cite{Hershey2014DeepUM, khobahi2019model, 8683876}. Note that the goal of waveform design for radar systems is to acquire the maximum amount of information from the desirable sources in the environment, where in fact, \emph{the transmit signal can be viewed as a medium that collects information}. In light of this, we employ the deep unfolding framework that aims to take the well-established iterative approaches, and design a deep architecture for waveform design in radar systems under different unimodular signal constraint, and boost the performance of the underlying inference optimization algorithm in terms of speed of convergence and effectiveness. 
\section{Radar Model--- and Signal Design Formulation}\label{sec:background}
Consider a radar system transmitting unimodular codes used to modulate a train of sub-pulses.
Let $\bs = [s_1~s_2~\cdots~s_N]^T \in \complexC^N$ denote the complex-valued probing sequence to be designed.
Under the assumptions of negligible intrapulse Doppler shift, and that the sampling is synchronized to the pulse rate, the received discrete-time base-band signal $\by$, after pulse compression and alignment with the current range cell of interest, can be modeled as follows \cite{4644058}:
\begingroup
\begin{align}\label{eq:1}
	\by = \bA^H\balpha + \bepsilon,\,\,\text{where}
\end{align}
\endgroup
\begingroup
\begin{align*} \label{eq:2}
\bA^H &= \begin{bmatrix}
s_1 & 0 & \cdots & 0 & s_N & s_{N-1} & \cdots & s_2 \\
s_2 & s_1 &  & \vdots & 0 & s_N &  & \vdots \\
\vdots & \vdots & \ddots & 0 & \vdots & \vdots & \ddots & s_N \\
s_N & s_{N-1} & \cdots & s_1 & 0 & 0 & \cdots & 0
\end{bmatrix}
\end{align*}
\endgroup
\begin{align}
\balpha &= [\alpha_0~\alpha_1~\cdots~\alpha_{N-1}~\alpha_{-N+1}~\cdots~\alpha_{-1}]^T \in \complexC^{2N-1}.
\end{align}
Here, the parameter $\alpha_0$ is the scattering coefficient of the current range cell, while $\{\alpha_k\}_{k\neq0}$ are that of the adjacent range cells contributing to the clutter, and $\bepsilon$ is the signal independent interference comprising of measurement noise as well as other disturbances such as jamming. The main goal in a radar system given the measurement model in \eqref{eq:1} is typically to design the probing signal $\bs$ such that it allows for an accurate recovery of the target scattering coefficient~$\alpha_0$. 

Note that, in model-based radar waveform design, the statistics of the interference and noise is usually assumed to be known, e.g., through stand-alone prescan procedures. Under such conditions, the waveform design boils down to constrained quadratic or fractional quadratic program as detailed in previous work \cite{4644058, soltanalian2014designing, 6472022, bose2019joint, bose2019waveform}. An example for waveform design criteria comes from the waveform's merit for resolvability along with clutter rejection. 
Namely, using a matched filter (MF) in the pulse compression stage, one can look for codes that maximize the following criterion:
\begin{eqnarray}\label{eq:f}
	f(\bs) = \frac{n(\bs)}{d(\bs)} \triangleq {\frac{|\bs^H\by|^2}{\sum_{k \neq 0} {|\bs^H\bJ_k\by|^2}}} = \frac{\bs^H \bA \bs}{\bs^H \bB \bs},
\end{eqnarray}
where $\bA = \by\by^H$,  $\bB = \sum_{k\neq 0}\bJ_k\bA\bJ_k^H$, and $\{\bJ_k\}$ are shift matrices satisfying $[\bJ_k]_{l,m} = [\bJ_{-k}^H]_{l,m} \triangleq \delta_{m-l-k}$, with $\delta_{(\cdot)}$ denotes the Kronecker delta function. Note that the above function can be interpreted as an oracle to a signal-to-interference-noise (SINR) ratio as the numerator represents the signal power and the denominator represents the combined interference and noise power after applying the matched filter. We further note that, to lower the implementation cost, it is desirable to use unimodular codes, i.e. $s_k=e^{j\phi_k},\,\,\phi_k \in [0, 2\pi), ~k \in \{1, \dots, N\},$ that attain the smallest peak-to-average ratio possible for transmit signals. As a result, one can consider the following fractional program in its general form for radar waveform design:
\begin{align}\label{eq:opt-decor}
    \max_{\bs}~~{\frac{\bs^H \bA \bs}{\bs^H \bB \bs}},\,\, \st~~ |s_k|=1, ~~ k \in \{1,\dots,N\}
\end{align}
Note that evaluating the objective function in \eqref{eq:opt-decor}, i.e. computing $f(\bs)$, only requires the knowledge of the transmit sequence $\bs$ and the observed vector $\by$ at the receiver. Nevertheless, solving the above optimization program is still NP-hard and very hard to tackle in general. In order to approximate the solution, one can resort to power method-like iterations specifically designed to tackle unimodular quadratic programs (UQPs) \cite{soltanalian2014designing}. In what follows, we reformulate the problem of \eqref{eq:opt-decor} as a UQP, and present the corresponding power method-like iterations that lays the groundwork for our proposed hybrid model-aware and data-driven \emph{adaptive} waveform design framework. 

Observe that both the numerator $n(\bs)$ and the denominator $d(\bs)$ of the objective function $f(\bs)$ are quadratic in $\bs$. Hence, in order to tackle the maximization of \eqref{eq:f} (or equivalently tackling \eqref{eq:opt-decor}) we resort to fractional programming techniques~\cite{dinkelbach,tofallis1998fractional}. Since $f(\bs)$, the SINR, is finite, we must have that $d(\bs) = \bs^H \bB \bs>0$. In addition, let $\bs_{\star}$ denote the current value of the code sequence $\bs$. Then, we define
\begin{align}\label{eq:17}
    e(\bs) &\triangleq n(\bs) - f(\bs_{\star})d(\bs),\\
    \bs_{\dagger} &= \underset{\bs}{\arg\max}\;\; e(\bs).\label{eq:18}
\end{align}
Henceforth, it can be easily verified by the virtue of  \eqref{eq:18} that $e(\bs_{\dagger})\geq e(\bs_{\star}) = 0$. As a result, we have that $e(\bs_{\dagger}) = n(\bs_{\dagger}) - f(\bs_{\star})d(\bs_{\dagger})\geq0$ implying that
\begin{eqnarray}\label{eq:19}
f(\bs_{\dagger})\geq f(\bs_{\star}),
\end{eqnarray}
as $d(\bs_{\dagger})>0$. In other words, we can argue that with respect to $\bs_{\star}$, the $\bs_{\dagger}$ increases the objective function $f(\bs)$. It is noteworthy to mention that for the criteria in \eqref{eq:19} to hold, it is sufficient for $\bs_{\dagger}$ to satisfy $e(\bs_{\dagger})\geq e(\bs_{\star})$ and that $\bs_{\dagger}$ shall not necessarily be the maximizer of $e(\bs)$.

For a given $\bs_{\star}$ maximizer of \eqref{eq:opt-decor} we have that:
\begin{align}
    e(\bs) = \bs^H\bA\bs - f(\bs_{\star})\left(\bs^H\bB\bs\right)=\bs^H\underbrace{\left(\bA - f(\bs_{\star})\bB\right)}_{\triangleq\tilde{\bchi}}\bs\nonumber
\end{align}
Now, in order to ensure that $\tilde{\bchi}$ is positive definite, one can perform a diagonal loading procedure by defining $\bchi \triangleq \tilde{\bchi} + \lambda\bI_N$, where $\lambda\geq \max\{0,-\lambda_{\min}(\tilde{\bchi})\}$. Next, the optimization problem of \eqref{eq:opt-decor} can be cast as the following UQP \cite{soltanalian2014designing}:
\begin{align}\label{eq:opt-decor2}
    \max_{\bs}~~{\bs^H\bchi\bs},\,\, \st~~ |s_k|=1, ~~ k \in \{1,\dots,N\}.
\end{align}

In order to efficiently tackle \eqref{eq:opt-decor2}, a set of \textit{power method-like iterations} (PMLI) were introduced in \cite{soltanalian2014designing, 6472022} that can be used to monotonically increase the objective value in \eqref{eq:opt-decor2} using the following nearest-vector problem:
\begin{align}\label{eq:powerR}
	\min_{\bs^{(n+1)}} ~{\left\| \bs^{(n+1)} - \bchi\bs^{(n)}\right\|_2},\;\;\st ~\left|s_k^{(n+1)}\right| = 1,\,\forall\,k.
\end{align}
The solution to \eqref{eq:powerR} can be computed analytically and is given as follows \cite{soltanalian2014designing,6472022}:
\begin{align}\label{eq:PMLIR}
    \bs^{(n+1)} = e^{j \arg(\bchi\bs^{(n)})}.
\end{align}
where $n$ denotes the internal iteration number, and $\bs^{(0)}$ is the current value of $\bs$. One can continue updating $\bs$ until convergence in the objective of \eqref{eq:opt-decor}, or for a fixed number of steps, say $L$. These iterations are already shown to provide a monotonic behavior of the quadratic objective (no matter what the signal constraints are), and subsume the well-known power method as a special case. Such a general approach to computationally efficient quadratic programming that can handle various signal constraints (many of which cause the problems to become NP-hard) opens new avenues in signal processing in low-cost scenarios. 

Note that, in many practical scenarios, one might not have access to the \textit{a priori} information about environmental parameters. In the following, we aim to devise a hybrid data-driven and model-based approach that allows us to jointly design adaptive transmit code sequences while learning these parameters given the fact that the environmental information are in fact embedded into the observed received signal $\by$. Namely, we propose a novel neural network structure for waveform design, \textbf{D}eep \textbf{E}volutionary \textbf{Co}gnitive \textbf{R}adar (DECoR), by considering the above power method-like iterations as a baseline algorithm for the design of a model-based deep neural network. In particular, we consider an over-parametrization of the power method-like iterations and unfold them onto the layers of a deep neural network. Each layer of the resulting network is designed such that it imitates one iteration of the form \eqref{eq:PMLIR}. Consequently, the resulting deep architecture is model-aware, uses the same non-linear operations as those in the power method, and hence, is interpretable (as opposed to general deep learning models). The structure yet allow us to utilize data-driven approaches to optimize the parameters of the network in an online learning manner---making the resulting network a great candidate for reliable adaptive waveform design in automotive radar applications.
\section{The DECoR Architecture for Signal Design}\label{sec:decor}

Consider the dynamics of a general fully connected deep neural network. Let $\tilde{g}_{\phi_i}$ be defined as
\begin{align}
\tilde{g}_{\phi_i}(\bz) = a(\bu), \,\, \text{where}\,\,\bu = \bW_i\bz,
\end{align}
where $\phi_i=\{\bW_i\}$ denotes the set of parameters of the function $g_{\phi_i}$, and $a(\cdot)$ denotes a non-linear activation function. Then, given an input $\bx_0$, the dynamics of a fully connected neural network with $L$ layers can be expressed as follows:
\begin{eqnarray}
    \bx_L = \mathcal{F}\left(\bx_0; \bUpsilon\right) = \tilde{g}_{\phi_{L-1}} \circ \tilde{g}_{\phi_{L-2}} \circ \dots \circ \tilde{g}_{\phi_0}(\bx_0),
\end{eqnarray}
where, for a general DNN, $\boldsymbol{\Upsilon} = \{\phi_i\}_{i=0}^{L-1}$ denotes the set of weight matrices $\bW_i$ for each layer. Now, consider the power method-like iterations of the form \eqref{eq:PMLIR}. The connection between the two becomes clear by paying attention to the fact that a fully connected DNN with an activation function defined as $a(\bx) = e^{j\arg(\bx)}$, and parameterized on a matrix $\bW$ (that is tied along the layers), boils down to performing $L$ iterations of the PMLI. Therefore, one can immediately see that a fully connected DNN with the specific choice of non-linear activation function given by the projection operator $\mathcal{S}(\bx) \triangleq e^{j\arg(\bx)}$ is an optimal architecture for waveform design with respect to the power method-like iterations extensively used in waveform design in various applications \cite{7472288, 8859282}. Hence, power method-like  iterations are  perfect candidates for unfolding into DNNs since they can be characterized by a linear step, followed by a possibly non-linear operation.

\subsection{The \textbf{D}eep \textbf{E}volutionary \textbf{Co}gnitive \textbf{R}adar Architecture}
\begin{figure*}[t]
	\centering
	\includegraphics[width=1\textwidth,draft=false]{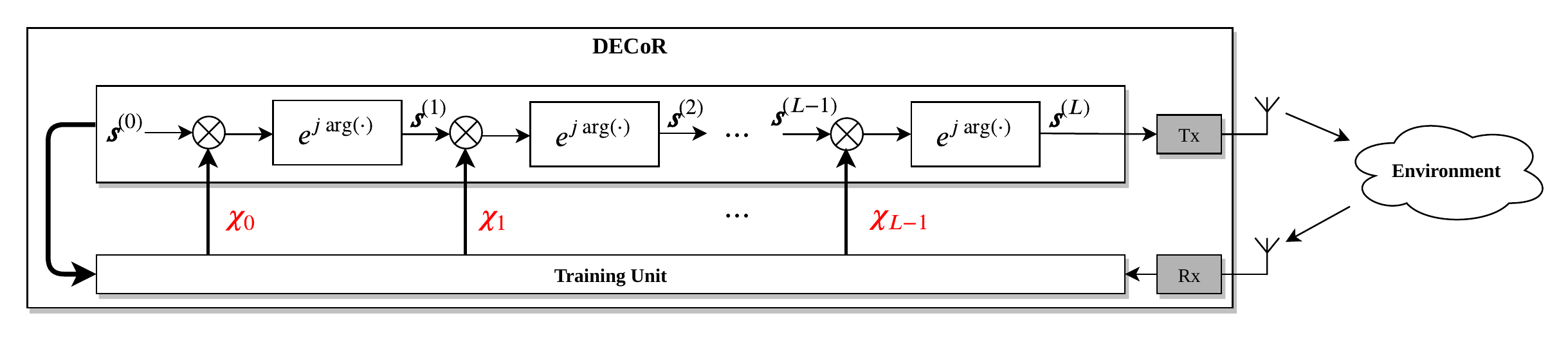}
	\caption{The proposed DECoR architecture for adaptive radar waveform design.} 
	\label{fig:decor}
\end{figure*}

The derivation begins by considering that in the vanilla PMLI algorithm, the matrix $\bchi$ is tied along all iterations. Hence, we enrich the PML iterations by introducing a weight matrix $\bchi_i$ \emph{per iteration} $i$. Note that in the original PMLI algorithm, the matrix $\bchi$ changes from one outer iteration to another. Hence, such an over-parameterization of the iterations results in a deep architecture that is faithful to the original model-based signal design method. Such an over-parametrization yields the following computation model for our proposed deep architecture (DECoR). Let us define $g_{\phi_i}$ as
\begin{align}
g_{\phi_i}(\bz) = \mathcal{S}(\bu), \;\;\; \text{where}~~\bu = \bchi_i\bz,
\end{align}
where $\phi_i=\{\bchi_i\}$ denotes the set of parameters of the function $g_{\phi_i}$, and recall that the non-linear activation function is defined as $\mathcal{S}(\bx) = e^{j\arg(\bx)}$ applied element-wise on the vector argument. Then, the dynamics of the proposed DECoR architecture with $L$ layers can be expressed as:
\begin{eqnarray}
    \bs_L = \mathcal{G}\left(\bs_0; \bOmega\right) = g_{\phi_{L-1}} \circ g_{\phi_{L-2}} \circ \dots \circ g_{\phi_0}(\bs_0),
\end{eqnarray}
where $\bs_{0}$ denotes some initial unimodular vector, and $\bOmega=\{\bchi_0,\dots,\bchi_{L-1}\}$ denotes the set of trainable parameters of the network. The block diagram of the proposed architecture is depicted in Fig. \ref{fig:decor}.

Our goal is to optimize the set of parameters $\bOmega$ of the proposed DECoR architecture using an online learning strategy that allows for fast adaptation to different environment. Intuitively, given the nature of the PML iterations, learning the parameters $\bOmega = \{\bW_l\}_{l=0}^{L-1}$ corresponds to learning the information corresponding to the signal dependent interference and environmental noise profile.
\subsection{The Proposed Online Learning Strategy}
In an automotive radar application, the environment might undergo drastic changes along different coherent processing intervals, and the noise and interference statistics might vary as a result. Hence, it is natural to consider an online learning strategy for training the proposed DECoR architecture. 

Let $\bOmega^{(t)}$ denote the set of parameters at time $t$. Then, the resulting code sequence given the set of parameters $\bOmega^{(t)}$ is simply given by the output of the last layer of the proposed DECoR architecture, i.e. $s^{(t)}_L = \mathcal{G}\left(\bs_0; \bOmega^{(t)}\right)$. We define the goal of our online training procedure as learning the set of parameters $\bOmega^{(t+1)}$ such that the resulted code sequence $\bs^{(t+1)} = \mathcal{G}\left(\bs_0; \bOmega^{(t+1)}\right)$ satisfies the following criterion:
\begin{eqnarray}\label{eq:training}
f(\bs^{(t+1)})\geq f(\bs^{(t)}).
\end{eqnarray}
Accordingly, we propose the following \emph{random walk-based} training strategy for optimizing the parameters of the proposed DECoR architecture in an online manner:
\noindent\rule{0.49\textwidth}{1pt}
\noindent $\bullet$ \underline{\textbf{\emph{Step 0}} (Initialization):} Choose an arbitrary unimodular transmit sequence $\bs_0\in\mathds{C}^N$, and set the training counter to $t=0$. Initialize the radius $\sigma$ of the search region to some positive constant $c$, and choose  $\delta\in(0,1]$. Further initialize the set of weight matrices $\bOmega^{(0)} = \{\bchi^{(0)}_i\}_{i=0}^{L-1}$ such that $\bchi_i^{(0)}\succ 0$, for $i\in\{0,\dots,L-1\}$.\\
$\bullet$ \underline{\textbf{\emph{Step 1}} (Random walk- generation):} For $l\in\{0,\dots,L-1\}$, generate $B$ random lower triangular matrices $\bL_l^0,\dots,\bL_l^{B-1}\sim\mathcal{C}\mathcal{N}(0,\sigma \bI)$, and form the set of Hermitian positive-definite search direction matrices $\small\bD_l^i = \bL_l^i\bL_l^{iH}$, for each layer $l$ and for $i\in\{0,\dots,B-1\}$, where $\bD^i_l\in\mathds{C}^{N\times N}$.\\
$\bullet$ \underline{\textbf{\emph{Step 2}} (Random walk- perturbation):} For $i\in\{0,\dots,B-1\}$, form the set of possible candidate updates for the current parameter space $\bOmega^{(t)}$ as $\bOmega_i^{(t)}=\{\bchi^{(t)}_0+\bD^i_0, 
\dots, \bchi^{(t)}_{L-1}+\bD^i_{L-1}\}$. Compute the corresponding $B$ unimodular codes $\bs^{(t)}_{L,i} = \calG(\bs_0;\bOmega^{(t)}_i)$ for $i\in\{0,\dots,B\}$ and form the set of training transmission codes as $\bS^{(t)} = \{\bs^{(t)}_{L,0}, \dots, \bs^{(t)}_{L,B-1}\}$.\\
$\bullet$ \underline{\textbf{\emph{Step 3}} (Collecting information):} Transmit the unimodular codes in the set $\bS^{(t)}$ and obtain the corresponding set of received signals $\bY = \{\by^{(t)}_0, \dots, \by^{(t)}_{B-1}\}$. Compute the function $f(\bs)$  for each transmit/receive pair $(\bs^{(t)}_{L,i}, \by^{(t)}_{i})$ and construct the set of objective values as $\mathcal{F}=\{f(\bs^{(t)}_{L,i})\}_{i=0}^{B-1}$.\\
$\bullet$ \underline{\textbf{\emph{Step 4}} (Optimizing the DECoR architecture):} Choose the current optimal parameter space using
\[
i_\star = \underset{i\in[B]}{\arg\max}\;\; f(\bs^{(t)}_{L,i}).
\]
Update the network parameters if $f(\bs^{(t)}_{L,i_\star})\geq f(\bs^{(t-1)}_{L})$ and set the search radius as $\sigma \leftarrow c$.
Otherwise, only update the search radius as $\sigma\leftarrow\delta\sigma$. Continue the online learning by going to Step~1. 

\noindent\rule{0.49\textwidth}{1pt}

The above proposed online learning strategy for the proposed DECoR architecture is an amalgamation of natural evolutionary optimization techniques and policy optimization in reinforcement learning. In particular, the increase in the objective function $f(\bs)$ can be seen as a task for an agent that is interacting with an unknown environment over the action space of $\bOmega$ and the corresponding unimodular code $\bs_L=\mathcal{G}(\bs_0,\bOmega)$. Note that the power method-like iterations and the model of the system impose a positive definite constraint on the weight matrices $\{\bchi_i\}_{i=0}^{L-1}$. In order to impose such a constraint in incrementally learning the parameters $\bOmega$, we initialize each $\bchi_i^{(0)}$ with some positive-definite matrix. We then perform a random walk in the cone of positive definite matrices by forming positive definite search direction matrices $\bD_l^i=\bL_l^i\bL_l^{iH}$. 
Such a training strategy results in a fast adaptation to the ever changing environment. Hence, the radar agent can continually perform the training on the fly.  
\section{Results and Concluding Remarks}\label{sec:numerical}
\begin{figure*}
	\centering
	\subfigure[]{\label{fig:fs-decor}	
		\centering
		\includegraphics[draft=false, width=0.45\textwidth]{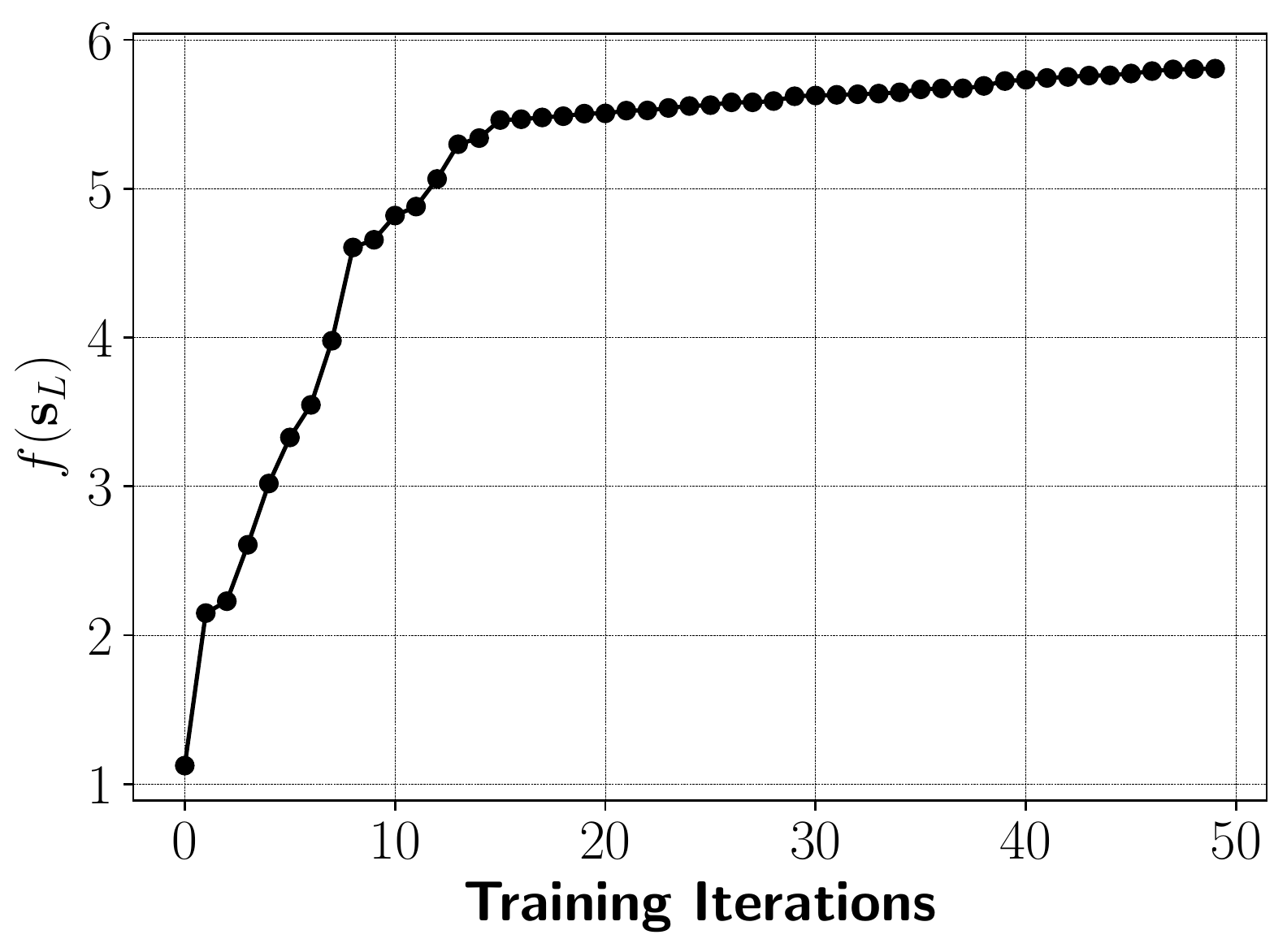}
	}
	\subfigure[]{\label{fig:mse-all}
		\centering
		\includegraphics[draft=false, width=0.48\textwidth]{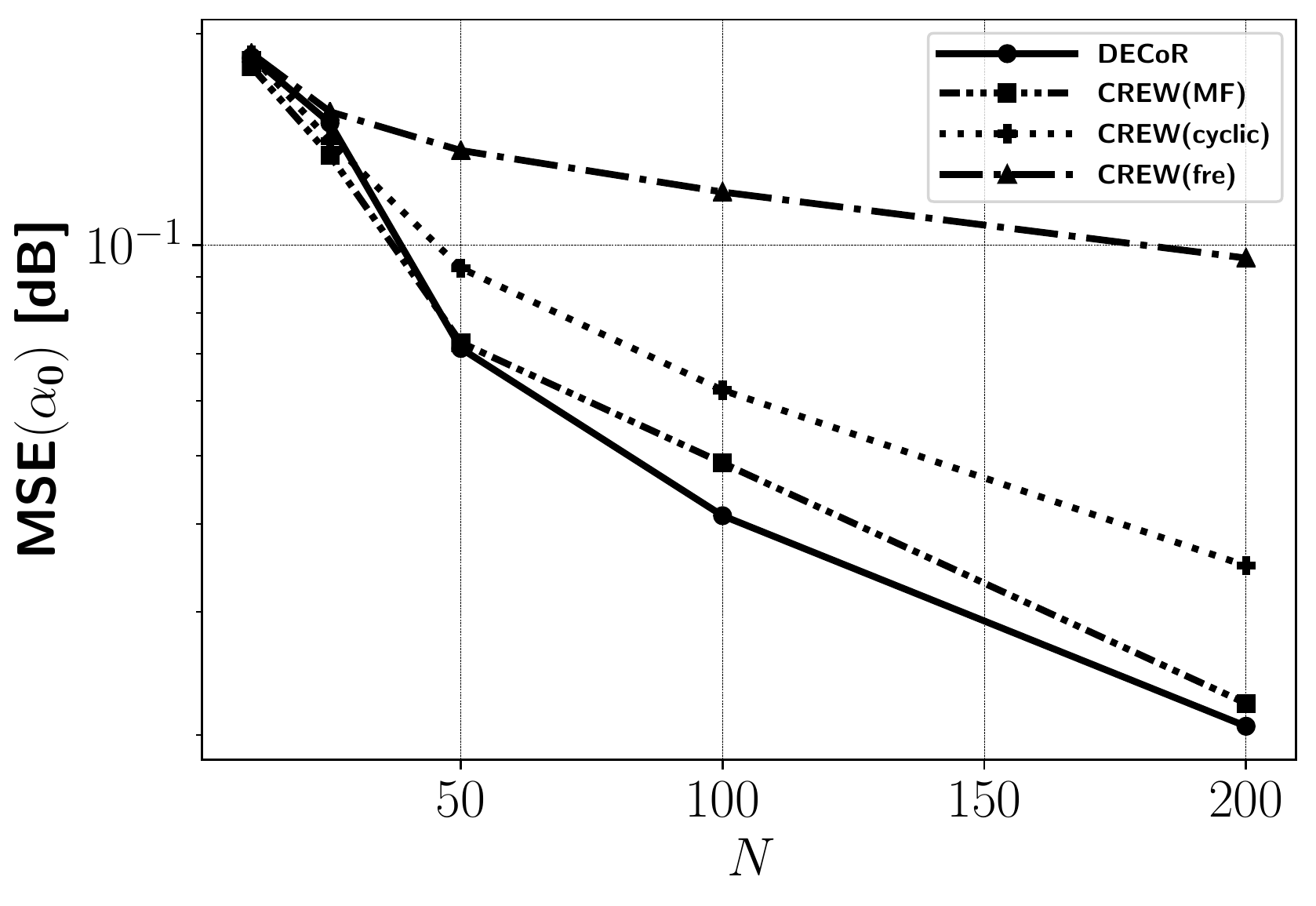}
	}
	\caption{Illustration of (a) the  objective  value $f(\bs_L)$ of the DECoR vs. training  iterations for a code length of $N=10$, and (b) MSE values obtained by the different design algorithms for code lengths $N \in \{10, 25, 50, 100, 200\}.$}
\end{figure*}
We begin by evaluating the performance and effectiveness of the proposed online learning strategy for optimizing the parameters of the DECoR architecture. For this experiment, we fix the total number of layers of the proposed DECoR architecture as $L=30$. Throughout the simulations, we assume an environment with dynamics described in \eqref{eq:1}, and with more details in \cite{4644058}, with clutter power  $\beta=1$, and a noise covariance of $\bGamma = \bI$. These information were not made available to the DECoR architecture and we only use them for data generation purposes. 

Fig. \ref{fig:fs-decor} demonstrates the objective value $f(\bs_L)$ in \eqref{eq:f} vs. training iterations, for a code length of $N=10$. It can be clearly seen that the proposed learning strategy and the corresponding DECoR architecture results in a \textit{monotonically} increasing objective value $f(\bs_L)$. Furthermore, note that the proposed learning algorithm optimizes the parameters of the proposed DECoR architecture very quickly. 
Next, we evaluate the performance of the presented hybrid model-based and data-driven architecture in terms of recovering the target coefficient $\alpha_0$. In particular, we compare the performance of our method (DECoR) in designing unimodular codes with two state-of-the-art model-based algorithms:
    (a) CREW(cyclic) \cite{6472022}, a cyclic optimization of the transmit sequence and the receive filter,
    (b) CREW(MF) \cite{6472022}, a version of CREW(cyclic) that uses a matched filter as the receive filter, and 
    (c) CREW(fre) \cite{4749273}, a frequency domain algorithm to jointly design transmit sequence and the receive filter. 
Fig.~\ref{fig:mse-all} illustrates the MSE of the estimated $\alpha_0$ vs. code lengths $N \in \{10, 25, 50, 100, 200\}$. For each $N$, we perform the optimization of DECoR architecture by allowing the radar agent to interact with the environment for $50$ training epochs. After the training is completed, we use the optimized architecture to generate the unimodular code sequence $\bs_L$ and use a MF to estimate~$\alpha_0$. We let the aforementioned algorithms to perform the code design until convergence, while the presented DECoR architecture has been only afforded $L=30$ layers (equivalent of $L$ iterations). 

It is evident that the proposed method significantly outperforms other state-of-the-art approaches. Although the DECoR framework does not have access to the statistics of the environmental parameters (as opposed to the other algorithms), it is able to learn them by exploiting the observed data from interaction with the environment.  


\bibliographystyle{IEEEtran}
\bibliography{bib-est}

\end{document}